\documentclass[%
reprint,
superscriptaddress,
nofootinbib,
 amsmath,amssymb,
 aps,
 physrev,
]{revtex4-1}

\usepackage[unicode=true,pdfusetitle,
 bookmarks=true,bookmarksnumbered=false,bookmarksopen=false,
 breaklinks=false,pdfborder={0 0 1},backref=false,colorlinks=true]{hyperref}
\hypersetup{
 urlcolor=blue, citecolor=blue}
\usepackage{amsmath}
\usepackage{tabularx}
\usepackage{footnote}

\usepackage[pdftex]{color}
\usepackage[pdftex]{graphicx}
\usepackage{dcolumn}
\usepackage{bm}

\usepackage{lineno}

\usepackage[caption=false]{subfig}

\usepackage[normalem]{ulem} 

\begin{document}

\preprint{APS/123-QED}

\title{$^{28}$Al Half-life Measurement and the negative mirror asymmetry between the $^{28}$Al($\beta^-$)$^{28m}$Si and $^{28}$P($\beta^+$)$^{28m}$Si decays}

\author{B.~Liu}
\email{bliu4@nd.edu}
\affiliation{ Department of Physics and Astronomy, University of Notre Dame, Notre Dame, IN 46556, USA}
\author{M.~Brodeur}
\affiliation{ Department of Physics and Astronomy, University of Notre Dame, Notre Dame, IN 46556, USA}
\author{D.W.~Bardayan}
\affiliation{ Department of Physics and Astronomy, University of Notre Dame, Notre Dame, IN 46556, USA}
\author{F.D.~Becchetti}
\affiliation{Department of Physics and Astronomy, University of Michigan, Ann Arbor, MI 48109, USA}
\author{C.~Boomershine}
\affiliation{ Department of Physics and Astronomy, University of Notre Dame, Notre Dame, IN 46556, USA}
\author{D.P.~Burdette}
\affiliation{ Department of Physics and Astronomy, University of Notre Dame, Notre Dame, IN 46556, USA}
\author{L.~Caves}
\affiliation{ Department of Physics and Astronomy, University of Notre Dame, Notre Dame, IN 46556, USA}
\author{O.~Olivas-Gomez}
\affiliation{ Department of Physics and Astronomy, University of Notre Dame, Notre Dame, IN 46556, USA}
\author{S.L.~Henderson}
\affiliation{ Department of Physics and Astronomy, University of Notre Dame, Notre Dame, IN 46556, USA}
\author{J.J.~Kolata}
\affiliation{ Department of Physics and Astronomy, University of Notre Dame, Notre Dame, IN 46556, USA}
\author{J.~Long}
\affiliation{ Department of Physics and Astronomy, University of Notre Dame, Notre Dame, IN 46556, USA}
\author{A.D.~Nelson}
\affiliation{ Department of Physics and Astronomy, University of Notre Dame, Notre Dame, IN 46556, USA}
\author{P.D.~O'Malley}
\affiliation{ Department of Physics and Astronomy, University of Notre Dame, Notre Dame, IN 46556, USA}
\author{A.~Pardo}
\affiliation{ Department of Physics and Astronomy, University of Notre Dame, Notre Dame, IN 46556, USA}
\author{R.~Zite}
\affiliation{ Department of Physics and Astronomy, University of Notre Dame, Notre Dame, IN 46556, USA}

\date{\today}

\begin{abstract}


In the past, the mirror asymmetry parameter has been proposed as a probing mechanism for the presence of beyond the Standard Model second-class currents in nuclear beta decay transitions. However, this was hindered by large uncertainties in the required nuclear structure correction terms. Recently, a new calculation of these corrections attempted, but could not fully explain the negative mirror asymmetry between the $^{28}$Al($\beta^-$)$^{28m}$Si and $^{28}$P($\beta^+$)$^{28m}$Si decays. To put the mirror asymmetry parameter on a more solid footing, the half-life of $^{28}$Al was measured for the first time using a radioactive ion beam at the Nuclear Science Laboratory of the University of Notre Dame. The new result, $t_{1/2}=$134.432(34)~s, is consistent with most of the past data except for one highly discrepant measurement. The new mirror asymmetry parameter of -3.5(10)$\%$ obtained still does not agree with nuclear structure calculations.


\end{abstract}


\maketitle



\section{\label{sec:intro}Introduction}


According to the Weinberg–Salam theory \cite{Weinberg1967,Salam1959,Glashow1959}, the electroweak interaction is purely vector and axial vector in nature. Atomic nuclei undergoing a beta decay transition obey this interaction. However, transitions of one nucleon type to another occur in the presence of surrounding nucleons that modify the interaction requiring the inclusion of the so-called recoil-order terms to the vector and axial-vector currents \cite{Holstein1974}. The most general Lorentz covariant form of the vector current requires the inclusion of weak magnetism and an induced scalar term. Likewise, the axial-vector current requires the addition of induced tensor and pseudo-scalar terms. Induced tensor and scalar currents are often referred to as second-class currents, and their presence would violate the Standard Model \cite{Grenacs1985}. Tests of the strong form of the conserved vector current hypothesis that directly relates the vector and weak magnetism form factors to their electromagnetic counterparts have so far been pointing towards the absence of an induced scalar term \cite{Grenacs1985}. 

Induced tensor currents can be probed by comparing the $ft$-value of the $\beta^+$ and $\beta^-$ transitions feeding a common isotope \cite{Wilkinson1970}. In the absence of such currents, the asymmetry parameter $\delta$ will take the form \cite{Xayavong2024}: 
\begin{equation}
\delta = \frac{ft^+}{ft^-} - 1 \approx (\delta_C^+ - \delta_C^-) - (\delta_{NS}^+ - \delta_{NS}^-) - (\delta_R^+ - \delta_R^-),
 \label{eq:asym}
\end{equation}
where the $\pm$ relates to $\beta^{\pm}$ transitions, $\delta_C$ is the isospin-symmetry breaking correction, and ($\delta_{NS}$, $\delta_{R}$) are the nuclear structure dependent and independent radiative corrections respectively. Unfortunately, searches for second-class currents from this method are hampered by the uncertainty on the above correction terms \cite{Wilkinson2000}. However, there has been an improvement on that front \cite{Xayavong2023} for some decays including the $^{28}$Al($\beta^-$)$^{28m}$Si and $^{28}$P($\beta^+$)$^{28m}$Si decays \cite{Xayavong2024}. The experimental mirror asymmetry of $-3.7(11)\%$, for these decays, is particularly interesting since it is negative, while most other values are positive. Ignoring both possible Beyond the Standard Model and radiative correction contributions, this would signify that there is a greater isospin symmetry breaking correction for the decay of the less proton-rich isotope than the more proton-rich isotope. However, previous shell model calculations yielded a rather different mirror asymmetry ranging from 0.6 to 11.5 depending on the interaction used \cite{Smirnova2003}. A recent shell model calculation that includes $p$-shell contributions and improvements of the isospin-mixing correction resulted in a positive asymmetry parameter of 3.66(325)$\%$ that still falls short of the experimental value.

\begin{figure}
\begin{center}
\includegraphics[trim = 0mm 0mm 0mm 0mm, clip, width=\linewidth]{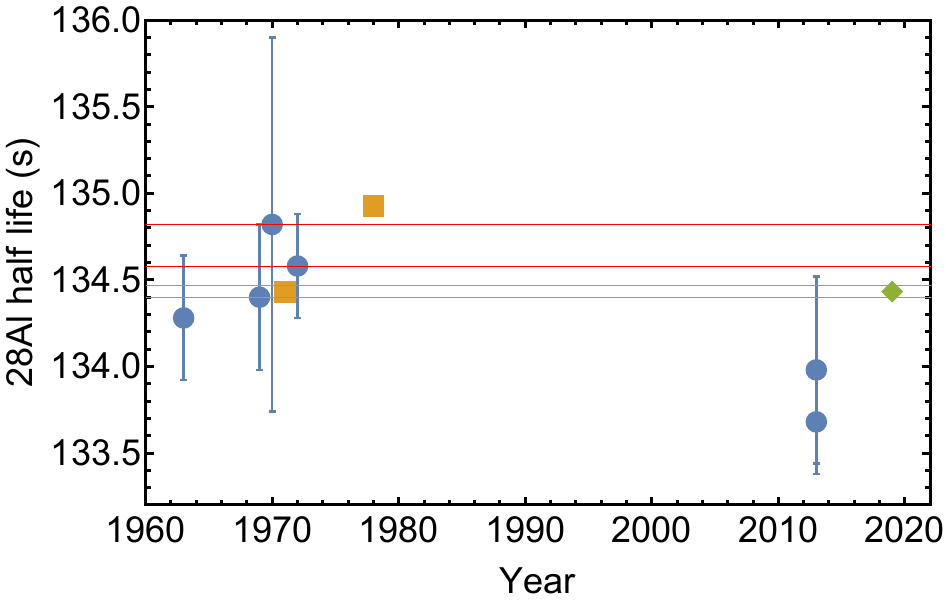}
\end{center}
\caption{Measurements adopted for the current world value for $^{28}$Al together with the 2013 measurement and the measurement result from this work (green). The two data points in 2013 represents the measurement from \cite{ZAHN201370} using two different types of corrections. The former precise measurements are denoted in orange while the others are denoted in blue. The green lines are the upper and lower limit of the $^{28}$Al half-life from this work. The red lines indicate the limits on the current world value.}
\label{worldvalue} 
\end{figure}

In light of these recent improvements in the theoretical corrections, it is imperative to ensure that all experimental data entering in the calculation of the asymmetry parameter ($Q$-values, branching ratios and half-lives) are on a solid foundation. Of these various quantities, the half-life of $^{28}$Al is particularly concerning, since its adopted value currently comes from a series of conflicting measurements. The current world value of the half-life of $^{28}$Al from the National Nuclear Data Center (NNDC)~\cite{Shamsuzzoha13} is 2.245(2) min. It consists of six measurements dating from 1963 to 1978 (see Fig.~\ref{worldvalue}) ~\cite{WEISS1963628,Wyttenbach1969,RYVES1970419,VanSchandevijl1971,NSR1972EM01,Becker1978}. Since 1978, this half-life has only been measured one additional time in 2013, and the result disagrees with the measurement of 1978 while agreeing with the rest of the measured values adopted for the world value~\cite{ZAHN201370}. Among the adopted measurements used to determine the world value of the $^{28}$Al half-life, the 1971 and 1978 measurements are both quoted to high precision, but are discrepant by 23 $\sigma$. This discrepancy increases the overall Birge ratio~\cite{Birge32} of all the adopted measurements to more than 10, resulting in the world value of the NNDC. This illustrates the lack of statistical consistency among these measurements and indicates the need for another independent measurement.

Furthermore, most of the current $^{28}$Al measurements used least squares fitting methods based on Gaussian statistics in their analysis. This fitting method has been shown to lead to a systematic bias in the results~\cite{ROBINSON197065,BAKER1984437}. Consequently, many evaluations, including \cite{Hardy15,Severijns08}, have rejected such results from their world values. Moreover, some early measurements \cite{WEISS1963628} did not consider dead-time corrections, while others might have been plagued with unidentified contaminants. All of these could potentially increase the inaccuracy and uncertainty in the current world value.

In an effort to remedy the situation, a precise half-life measurement of $^{28}$Al was performed using the $\beta$ counting station~\cite{Brodeur06} at the Nuclear Science Laboratory (NSL) of the University of Notre Dame. This was the first measurement of the half-life of $^{28}$Al performed using a radioactive ion beam, and the result was analyzed using a Poisson maximum likelihood fitting-based approach. The measurement result together with a new world value will be presented below.

 
\section{\label{sec:exp}Experimental Methods}
A stable beam of $^{27}$Al was created by extracting $^{27}$Al-$^{16}$O$^-$ molecular ions from the cathode wheel of the NSL-cesium sputtering ion source (SNICS) and then sent to the NSL FN Tandem Van de Graaff accelerator operated with a terminal voltage of 8.68 MV. A 3 $\mu$m carbon foil in the terminal broke the molecular ion and produced a distribution of $^{27}$Al charge states. The analyzing magnet downstream of the accelerator selected the 7+ charge state of $^{27}$Al, which had an energy of 69 MeV. The primary beam bombarded a deuterium gas target, producing $^{28}$Al through the (d, p) reaction. The secondary beam was then selected via the two superconducting solenoids of the \emph{TwinSol} radioactive beam facility~\cite{BECCHETTI2003377} before it arrived at the beta counting station. 

\begin{figure}
\begin{center}
\includegraphics[trim = 0mm 3mm 0mm 2mm, clip,width=7cm]{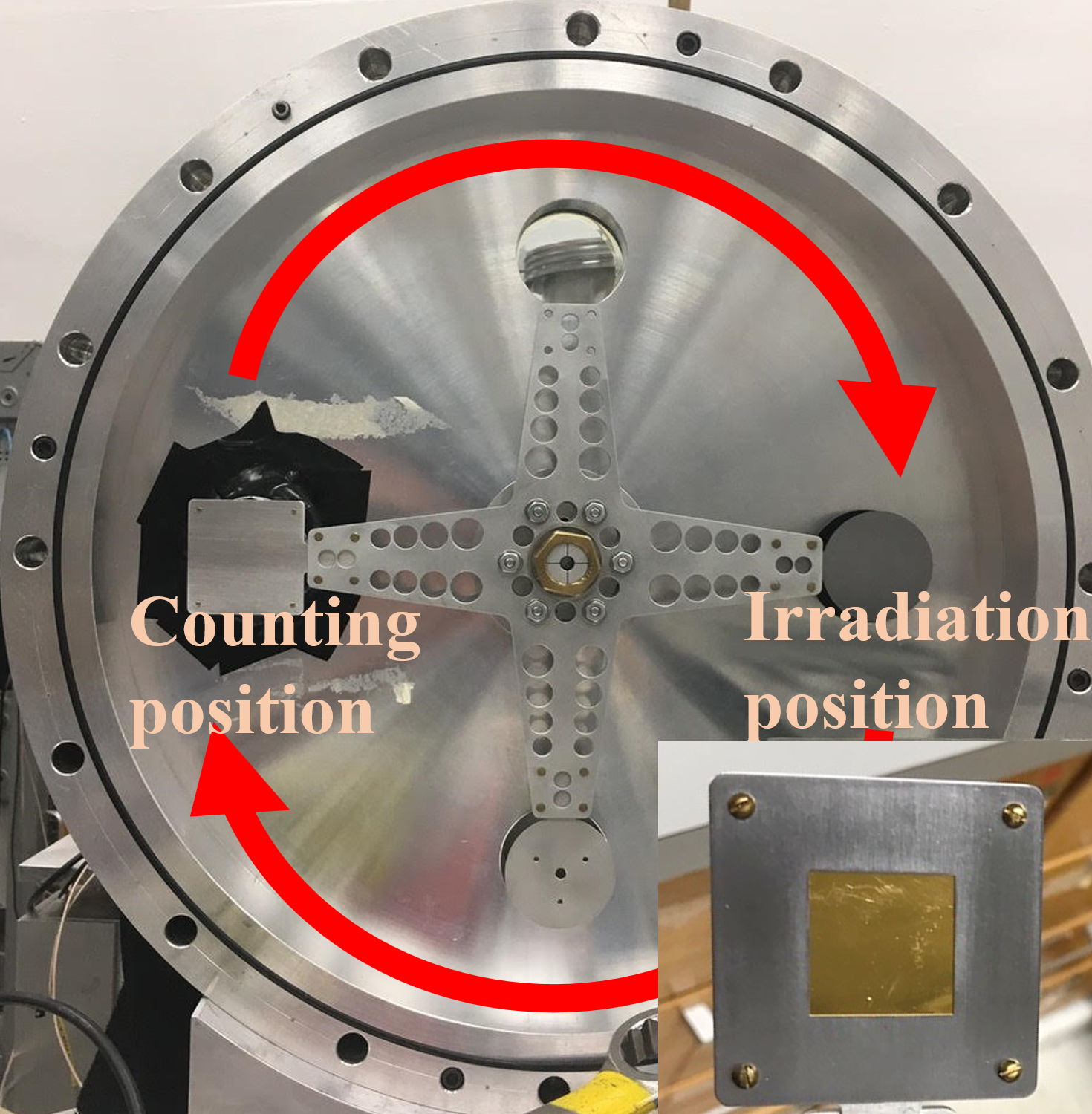}
\end{center}
\caption{\label{counter} The beta counting station at the NSL with the irradiation position and counting position marked. Inset: the implantation gold foil mounted on the paddle on the beam-facing side.}
\end{figure}

Fig.~\ref{counter} shows a photo of the beta counting station. The implantation paddle is made of aluminum with a 0.25 mm thick high-purity (99.999$\%$) gold foil mounted on it and is surrounded by a tantalum frame. For a typical measurement, the radioactive beam hits the gold at the radiation position for a certain time, and then the paddle rotates 180 degrees to the counting position. In the counting position, there is a 1 mm thick plastic scintillator connected to a photomultiplier tube (PMT) via a cemented light guide. When radioactive elements implanted in the gold foil decay and emit $\beta$ particles, these $\beta$ particles hit the plastic scintillator causing it to emit light. This signal is then sent to the PMT and recorded as events. After a fixed counting time, the paddle rotates back to the irradiation position to complete the cycle. Note that during the counting time the primary beam was deflected off the production target by an electrostatic steerer upstream of the FN accelerator, preventing any beam from impinging on the counting station. 

At a given setting of PMT voltage and discriminator threshold, usually 2 or 3 cycles were conducted sequentially forming a single run. The PMT voltage and discriminator threshold were varied between runs. During this measurement, 10 runs for a total of 27 cycles were recorded. The irradiation time and the counting time for all the 27 cycles were 400~s and 3250~s, respectively. PMT voltages and discriminator thresholds were varied from run to run to investigate possible systematic effects.


\section{\label{sec:data}Data analysis}

The data were first analyzed cycle-by-cycle to check if any data set was actually from an incomplete cycle, and also to probe for potential systematic uncertainties. Then a sum fit was performed as described in \cite{KOSLOWSKY1997289} to calculate the final half-life result. In order to be later summed together, all data sets were trimmed to a length of 3200 s for analysis. Instead of using least-squares fitting based on Gaussian statistics for the analysis, a Poisson maximum likelihood fitting was carried out iteratively by performing least-squares fittings \cite{KOSLOWSKY1997289,Fowler2014} using a Levenberg–Marquardt algorithm. Note that data sets were always corrected for dead time before fitting.

\subsection{\label{sec:deadtime}Dead time determination}

The data acquisition dead time was determined by two independent methods. In the first method, the dead time was extracted directly from the $^{28}$Al data for the 27 cycles. The total dead time for a given cycle is the difference between the clock time and the live time of the clock for the whole cycle. The dead time per event is this total dead time divided by the total number of counts. Taking an average of all the cycles, we obtained a dead time per event of 56.41(14)  $\mu$s.

The other method is the source-pulser method described in~\cite{Baerg_1965}. Through this method, the dead time per event was measured to be 56.47(11) $\mu$s~\cite{Long20} after our $^{29}$P measurement~\cite{Long20} and before this $^{28}$Al measurement.

The dead-time values of both methods agree with each other. Thus, the final dead time per event adopted for the whole analysis is the weighted average of the two independent and consistent results, 56.44(9) $\mu$s. This value was used for dead-time corrections in the analysis.

\subsection{\label{sec:cyclefit} Cycle-by-cycle analysis}

\begin{figure}
 \centering
    \centering
    \includegraphics[trim =0mm 0mm 0mm 0mm, clip,width=\linewidth]{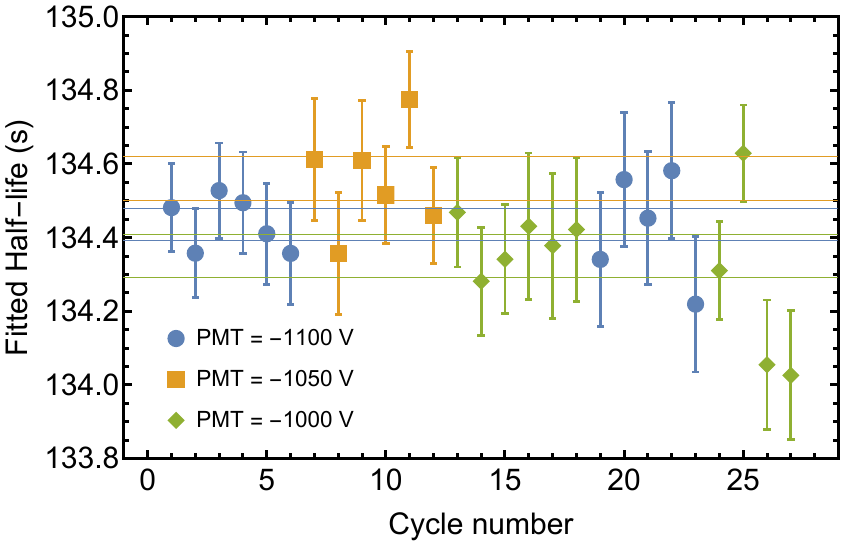}

     \hfill
         \centering
  
    \includegraphics[trim = 0mm 0mm 0mm 0mm, clip,width=\linewidth]{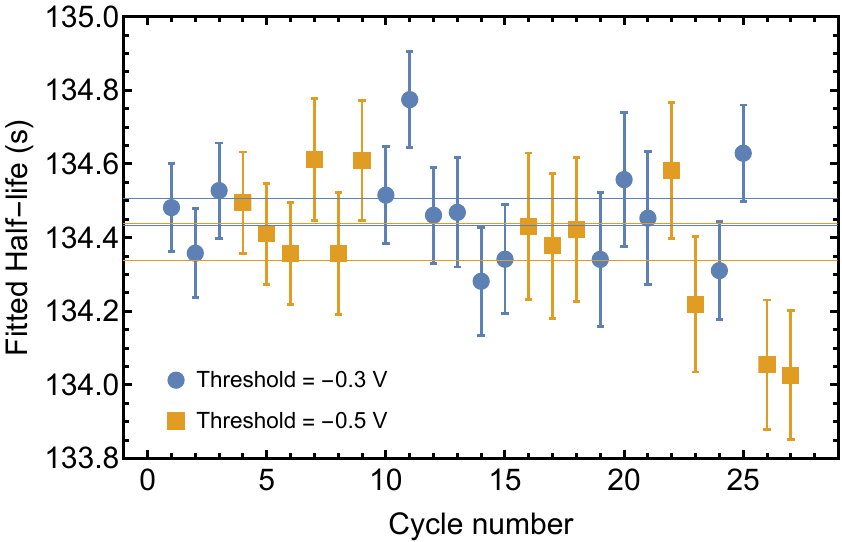}

\caption{\label{cyclenum}  Fitted half-lives for each cycle with PMT setting (top) and discriminator threshold setting (bottom) indicated. The colored lines indicate the uncertainty bands of the weighted average.}

\end{figure}

In the cycle-by-cycle analysis, the dead-time corrected data of each cycle were independently fitted to the integral of the exponential decay rate function given by Eq.~\ref{eq:decay1} 
\begin{equation}
r(t)=r_0 \, e^ {-\mathrm{ln}(2)(t/t_{1/2})} + b \label{eq:decay1}
\end{equation}
where $t_{1/2}$ is the half-life of $^{28}$Al, $r_0$ is the initial activity and $b$ represents the background. None of the 27 cycles were excluded because of low counts or abnormal decay patterns. The fitting result of all the cycles is shown in Fig.~\ref{cyclenum}. The PMT voltage was varyingly set to -1000, -1050, and -1100 V and the discriminator threshold was varyingly set to -0.3 and -0.5~V. As the bottom panel indicates, the fitting results of cycles with -0.3 and -0.5~V discriminator threshold are consistent, while the top panel indicates that the -1050 V PMT bias data results in a somewhat longer half-life than the -1000 V and the -1100 V data.

\begin{figure}
 \centering
    \centering

\includegraphics[trim =0mm 0mm 0mm 0mm, clip,width=\linewidth]{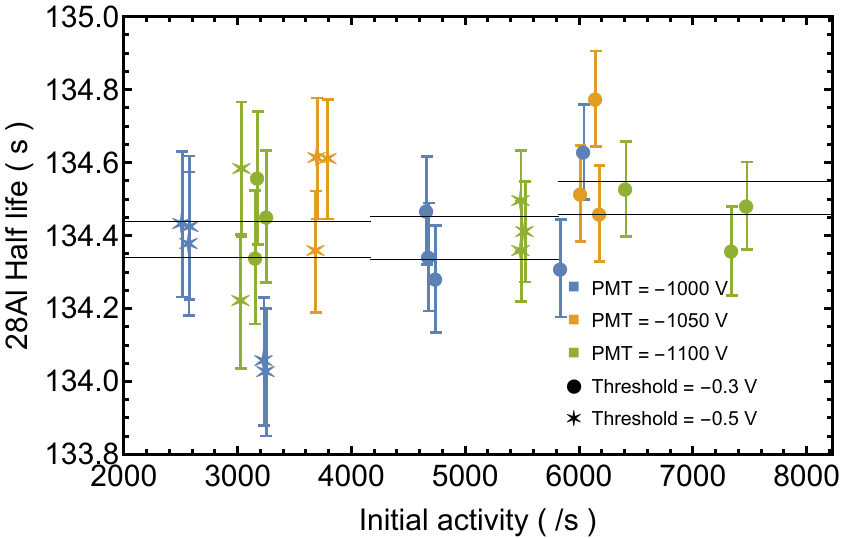}

     \hfill
  \centering
  
  \includegraphics[trim = 0mm 0mm 0mm 0mm, clip,width=\linewidth]{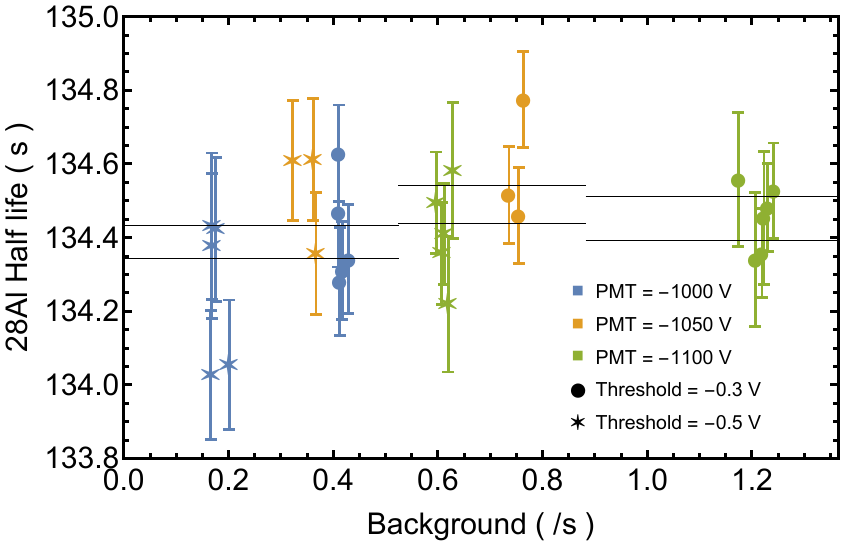}

\caption{\label{initialandbackground}  Fitted half-lives of 27 cycles as a function of initial activity (top panel) or background (bottom panel). PMT settings and discriminator threshold settings are indicated by different colors and shapes. The black lines represent the mean half-lives of the trisection between the lowest and highest initial activity or background value.}

\end{figure}

To further investigate for systematic uncertainties, the fitting results are reorganized in Fig.~\ref{initialandbackground} as a function of the initial activity and background. As shown, there seems to be no systematic correlation between the half-life and the initial activity or the background. 

\subsection{\label{sec:groupfit} Subgroup fit}

\begin{figure}
\begin{center}
\includegraphics[trim = 0mm 0mm 0mm 0mm, clip,width=\linewidth]{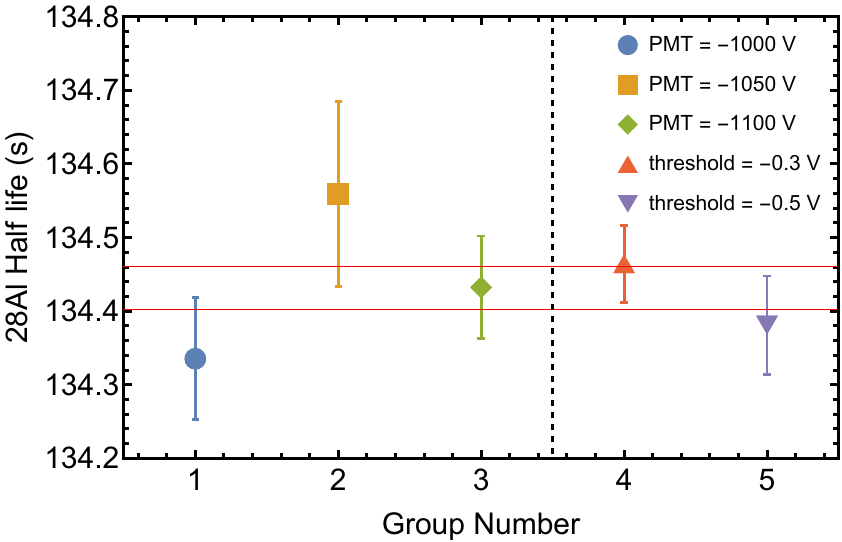}
\end{center}
\caption{\label{subgroup} Fit results of each subgroup of the cycles sharing the same PMT or discriminator threshold setting. The red lines represents the sum fit result in Sec.~\ref{sec:sumfit}.}
\end{figure}

As described in Sec.~\ref{sec:cyclefit}, the cycles with 1050 V PMT voltage tend to give a relatively higher mean half-life. To evaluate this effect, we subgrouped the cycles with the same PMT or discriminator threshold setting and fit the sum of all cycles in each group. The fit results are shown in Fig.~\ref{subgroup}. The Birge ratio of the three PMT subgroups is 1.1(3) while the Birge ratio of the two discriminator subgroups is 1.0(3). The Birge ratios are both close to 1, which means that the data fluctuation in each subgroup can be accounted for by statistical fluctuation.

\subsection{\label{sec:sumfit}Sum fit}
A sum fit of all cycles was performed as described in \cite{KOSLOWSKY1997289} was performed. All data from the 27 cycles were summed, dead-time corrected, and rebinned into 500 bins. The determination of bin number is discussed in Sec.~\ref{sec:uncertainties}. 

\begin{figure}
\begin{center}
\includegraphics[trim = 10mm 0mm 0mm 0mm, clip,width=\linewidth]{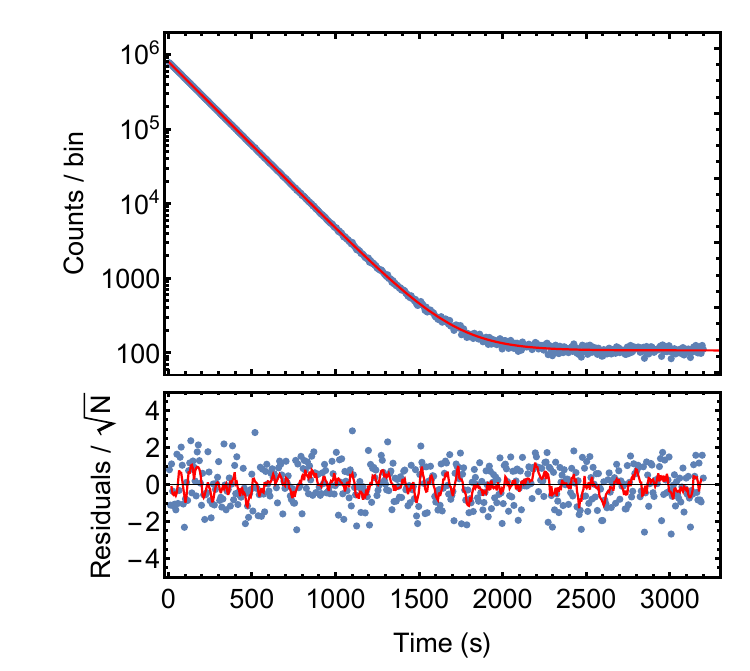}
\end{center}
\caption{\label{sumfit} Summed data and the sum fit curve (top). The residual of the fit divided by the square root of the number of counts in the given bin, and a 5-point moving average, are shown (bottom).}
\end{figure}

The summed data were fitted using Eq.~\ref{eq:decay1}, and are plotted with the fitting curve in the top panel in Fig.~\ref{sumfit}. The bottom panel shows the residuals, which are the differences between the summed data and the fitting curve, divided by the square root of the number of counts in the given bin along with a 5-point moving average. The residuals fluctuate around 0 with no obvious up- or down-trend. The sum fit gives a half-life of 134.432(29) s with a reduced $\chi ^2$ of 1.00.

\begin{figure}
\begin{center}
\includegraphics[trim = 0mm 0mm 0mm 0mm, clip,width=\linewidth]{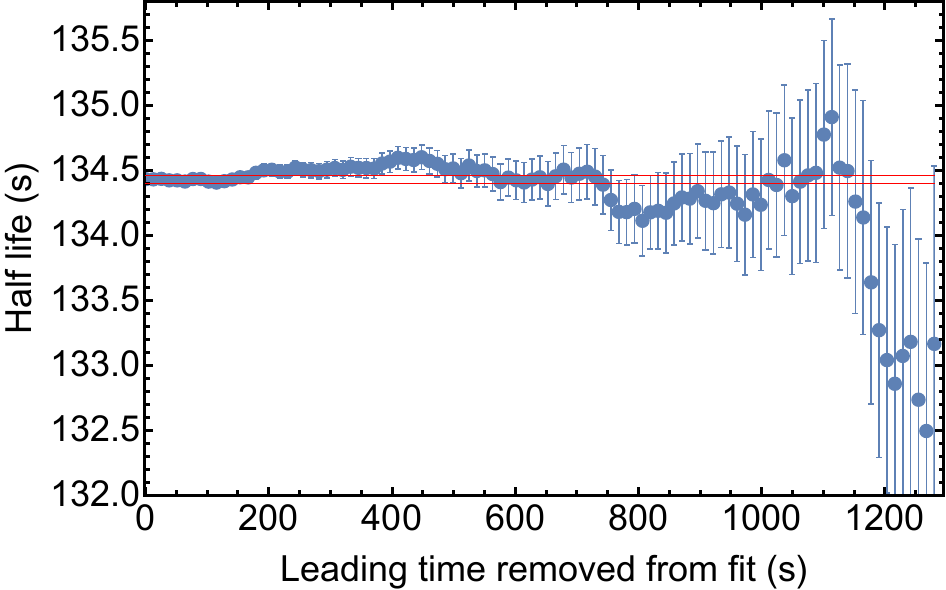}
\end{center}
\caption{\label{channel} The fitted half-life as a function of leading time removed from the beginning. The double red line represent the uncertainty on the fit of the complete decay curve.}
\end{figure}

To investigate potential contaminants, a channel removal analysis was performed. The sum fit was repeated as more and more of the leading bins were removed. Fig.~\ref{channel} shows the fitted half-life as a function of the removed leading time. As the plot shows, there is a bump around 400 s, which could either indicate potential contaminants in the beam or could be statistical in nature.

To investigate this indication, the summed data were further fitted using Eq.~\ref{eq:decay2} and Eq.~\ref{eq:decay3}. 

\begin{equation}
r(t)=r_0 \, (e^ {-\mathrm{ln}(2)(t/t_{1/2})} + R \, e^ {-\mathrm{ln}(2)(t/t_{1/2}^{'})}) + b \label{eq:decay2}
\end{equation}

\begin{equation}
r(t)=r_0 \, e^ {-\mathrm{ln}(2)(-t/t_{1/2})} + mt + b \label{eq:decay3}
\end{equation}

In Eq.~\ref{eq:decay2}, a contaminant with half-life $t_{1/2}^{'}$ is assumed to decay with an initial activity ratio $R$ with respect to $^{28}$Al. Every combination of the ratio 1, 0.1, 0.01, 0.001, 0.0001, 0.00001 and the contaminant half-life from 1, 10, 100, 1000 to 10000 s were tested and no contaminants were found within our level of statistical uncertainty. In Eq.~\ref{eq:decay3}, a potential long-lived contaminant was represented by the term $mt$. $m$ was tested from -1 to -$10^{-9}$. Again, long-lived contaminants were found to be absent within our level of statistical uncertainty. 

Since no contaminants could be identified, the sum fit result of Eq.~\ref{eq:decay1} is taken as our result. However, in Sec.~\ref{sec:groupfit}, the Birge ratio of the PMT subgroups is 1.1(3). Although this Birge ratio is consistent with one within uncertainty, to be conservative in estimating a potential underestimation in the fitting uncertainty, we increased the uncertainty directly from the fitting by the 1.1(3) Birge ratio, resulting in a half-life of 134.432(31) s.

\subsection{\label{sec:uncertainties}Systematic uncertainties}

\begin{table}
\caption{\label{tab:uncertainty}
Systematic uncertainty from dead time, clock time, binning effect and fitting.
}
\begin{ruledtabular}
\begin{tabular}{cc}
\textrm{Source}&
\textrm{Uncertainty (s)}\\
\colrule
 Dead time & 0.014 \\
 Clock time & 0.001 \\
 Binning effect & 0.001\\
 Fitting & 0.031\\
 \colrule
 Total uncertainty & 0.034\\
\end{tabular}
\end{ruledtabular}
\end{table}

\subsubsection{\label{sec:deadtime}Dead time}
To study the uncertainty of the half-life that stems from dead time, the upper limit and the lower limit of the dead time per event 56.44(9)~$\mu$s were used as dead time to perform the sum fit separately. Half of the difference between the two fitted half-lives results in 0.014~s and that is taken as the uncertainty from the dead time. As shown in Table~\ref{tab:uncertainty}, the uncertainty from dead time together with the uncertainty directly from the fitting are the two main sources of the total uncertainty.

\subsubsection{\label{sec:clock}Clock time}
The clock frequency was measured using a Teledyne Lecroy 500-MHz oscilloscope as 99.9996(10) Hz. Following the same routine as for the dead time, the upper and lower limits of the clock frequency were used to perform the sum fit, and half of the difference of the half-life results (0.001) s is taken to be the systematic uncertainty due to the clock time.

\subsubsection{\label{sec:binbing}Binning effect}

The number of bins has an effect on the accuracy of the fit. Too many bins would result in empty bins, which could introduce bias into the fit result, while too few bins will also decrease the accuracy~\cite{Long17}. Table~\ref{tab:bin} shows the fitted half-life with different bin numbers. As shown, 500 to 4000 bins give us the same half-life value within our uncertainty. 500 bins were used for the sum fit while 250 bins were used for the cycle-by-cycle analysis since a smaller bin number provides less possibility to contain empty bins. Half of the difference from the 250-bin and 1000-bin fitting results in 0.001 s and is taken as the uncertainty for our binning.

\begin{table}
\caption{\label{tab:bin}
Binning effect on the half-life fitting result.
}
\begin{ruledtabular}
\begin{tabular}{cc}
\textrm{Number of bins}&
\textrm{Half-life (s)}\\
\colrule
100 & 134.451(29) \\
200 & 134.437(29) \\
250 & 134.434(29) \\
500 & 134.432(29) \\
1000 & 134.432(29) \\
2000 & 134.432(29) \\
4000 & 134.432(29) \\
8000 & 134.434(29) \\
\end{tabular}
\end{ruledtabular}
\end{table}

All the above systematic uncertainties and the uncertainty of the fitting are listed in Table~\ref{tab:uncertainty}. All uncertainties were added in quadrature, resulting in a half-life of $^{28}$Al of 134.432(34) s.

\section{\label{sec:world}New world value}

\begin{table}
\caption{\label{tab:allhalflives}
Past $^{28}$Al half-lives included in the last NNDC evaluation \cite{Shamsuzzoha13} as well as a 2013 measurement \cite{ZAHN201370} and the measurement from this work.
}
\begin{ruledtabular}
\begin{tabular}{ccc}
\textrm{Year}&
\textrm{Half-life (s)}&
\textrm{Ref.}\\
\colrule
 1963 & 134.28(36) & \cite{WEISS1963628} \\
 1969 & 134.40(42) & \cite{Wyttenbach1969} \\
 1970 & 134.8(11) & \cite{RYVES1970419} \\
 1971 & 134.430(18)\footnote{The original uncertainty of 0.003 minutes from \cite{VanSchandevijl1971} was used rather than the 0.008 minutes uncertainty appearing in \cite{Shamsuzzoha13}.} & \cite{VanSchandevijl1971} \\
 1972 & 134.58(30) & \cite{NSR1972EM01} \\
 1978 & 134.930(9) & \cite{Becker1978} \\
 2013 & 133.98(54)\footnote{using the non-paralyzable correction}/133.68(30) \footnote{using the secondary correction} & \cite{ZAHN201370} \\
 2021 & 134.432(34) & this work \\
\end{tabular}
\end{ruledtabular}
\end{table}

With our new $^{28}$Al half-life value, we proceeded with an evaluation of the world data starting with the measurements that are in the NNDC value of \cite{Shamsuzzoha13}, and including the additional 2013 measurement from \cite{ZAHN201370}. All of these values are shown in Fig.~\ref{worldvalue} and Table~\ref{tab:allhalflives}. Following the Particle Data Group procedure \cite{Ber12}, we first exclude measurements that have an uncertainty that is more than ten times larger than the most precise measurement. The remaining measurements are the 1971 \cite{VanSchandevijl1971} and 1978 \cite{Becker1978} measurements and the measurement from this work. Our measurement is in excellent agreement with the 1971 value, but disagrees with the 1978 value by more than 14 standard deviations.

In their evaluation, \cite{Shamsuzzoha13} dealt with the discrepancy by introducing an uncertainty of 0.1$\%$ into the 1978 data. If we take a weighted average of all data included in \cite{Shamsuzzoha13} and using the published uncertainty of \cite{VanSchandevijl1971} (see Table~\ref{tab:allhalflives} footnote), we obtain a half-life of 134.774(10) s and a Birge ratio of 10.3. If we then follow the procedure of the Particle Data Group \cite{Ber12} by inflating the uncertainty on the average by the Birge ratio, we obtain a half-life of 134.77(10) s, which is close to the value of \cite{Shamsuzzoha13} and also has an  uncertainty of 0.1~s. If we follow the same procedure this time including our new measurement, the average half-life becomes 134.748(10)~s (without adjustment by the Birge ratio) and the Birge ratio increases to 17.7, which makes the discrepancy even more evident.

The 1971 publication \cite{VanSchandevijl1971} focused on the $^{28}$Al half-life measurement and presented it in sufficient detail to appreciate that due diligence was taken in performing the measurement, although least squares fitting was used in the analysis, which can sometimes introduce a bias in the results. The 1978 publication \cite{Becker1978} on the other hand focused on the half-life measurement of $^{14}$O. Although the measurement and analysis of $^{14}$O is presented in great detail, $^{28}$Al only appears in a table with other isotopes: $^{20}$F (not included in the NNDC evaluation of $^{20}$F), $^{25}$Na, and $^{42}$Sc. In that table, $^{28}$Al is by far the most precise value reported, with a relative uncertainty of 7$\times$10$^{-5}$, which is 16 times more precise than their determination of $^{14}$O, which was the main focus of the paper. Furthermore, the authors mention in the text that ``Half-life results for several other nuclides have been obtained, although we have not given these measurements the detailed care accorded to $^{14}$O''. Hence, we believe that the reported $^{28}$Al uncertainty in \cite{Becker1978} is too small and there also appear to be possible misprints in their table of results. Due to the suspicious nature of this result, we recommend rejecting the \cite{Becker1978} $^{28}$Al half-life measurement from future evaluations. 
If \cite{Becker1978} is excluded from the evaluation of the world average becomes 134.430(16)~s.

\section{\label{sec:world}Asymmetry parameter}

\begin{table}
\caption{\label{tab:asymmetry}
Quantities entering in the calculation of the asymmetry parameter for the $^{28}$Al($\beta^-$)$^{28m}$Si and $^{28}$P($\beta^+$)$^{28m}$Si decays. Both decays feed a 1779.030(11)~keV excited state in $^{28}$Si. Log$ft$-values were calculated using the code BetaShape \cite{Turkat2023}.}
\begin{ruledtabular}
\begin{tabular}{ccc}
\textrm{Parameter}&
\textrm{$^{28}$Al}&
\textrm{$^{28}$P}\\
\colrule
 $Q_{\beta^{\pm}}$ & 4642.078(49) keV & 14344.9(11) keV \\
 $P_{EC}$ & N/A  & 0.00237(7)$\%$ \\
 $t_{1/2}$ & 134.430(16) s & 270.3(5) ms \\
 BR & 99.99(1)$\%$ & 69.1(7)$\%$ \\
 log$ft$ & 4.87106(8)  & 4.8556(45)  \\
\end{tabular}
\end{ruledtabular}
\end{table}

With the suggested $^{28}$Al half-life of 134.430(16)~s, and the various parameters in Table~\ref{tab:asymmetry} (all from NNDC \cite{Shamsuzzoha13}), we calculated the log$ft$-value for the $\beta^-$ decay of $^{28}$Al and the $\beta^+$ decay of $^{28}$P into the $T=1$, $J_{\pi}=2^+$, $E_x$ = 1779.030(11)~keV excited state state in $^{28}$Si using the code BetaShape \cite{Turkat2023}.
The 2020 Atomic Mass Evaluation \cite{AME2020} $Q_{\beta}$ values were used in these calculations. 
\begin{figure}
\begin{center}
\includegraphics[trim = 0mm 0mm 0mm 0mm, clip,width=\linewidth]{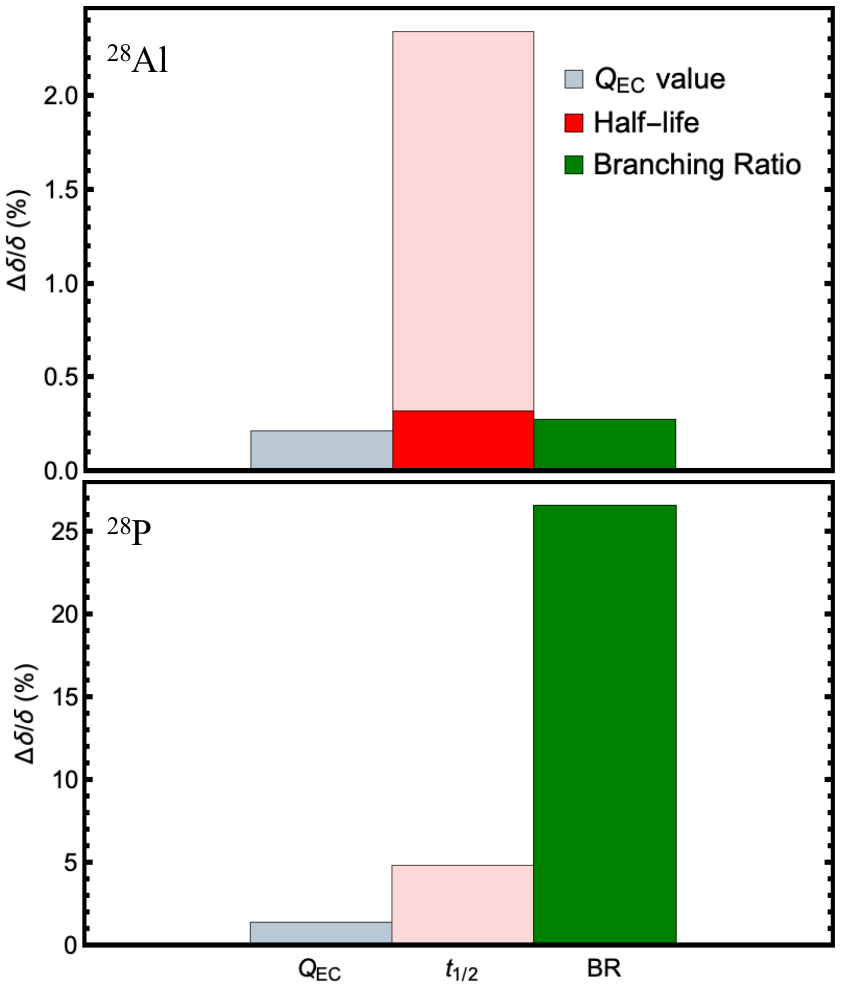}
\end{center}
\caption{\label{ftvalue} Relative changes in the asymmetry parameter $\Delta \delta/\delta$ due to the uncertainty of the experimental quantities entering in its determination. The darker red represents the new $^{28}$Al half-life.}
\end{figure}
Then, using these log$ft$-values and the left-hand side of Eq.~\ref{eq:asym}, we calculated a new asymmetry parameter $\delta$
for the $^{28}$Al($\beta^-$)$^{28m}$Si and $^{28}$P($\beta^+$)$^{28m}$Si decays. The new value $\delta$ = -3.5(10)$\%$ is slightly less negative than the previous value of $\delta$ = -3.7(11)$\%$.

Figure~\ref{ftvalue} shows the relative uncertainty in the asymmetry parameter $\Delta \delta/\delta$ due to the uncertainty of the individual experimental quantities presented in Table~\ref{tab:asymmetry}. Before the present measurement of the half-life of $^{28}$Al, the relative uncertainty due to this experimental quantity was the third largest, only exceeded by the branching ratio and half-life of $^{28}$P decay.

As indicated in Figure~\ref{ftvalue}, with the new $^{28}$Al half-life, the main sources of uncertainty in the determination of the asymmetry parameter are all related to experimental quantities entering in the computation of the log$ft$-value of $^{28}$P.

The largest source of uncertainty is from the branching ratio of the $^{28}$P $\beta^+$ decay to the $T=1$ state at $E_x$ = 1779.030(11)~keV. That branching ratio was measured in 1982 \cite{Warburton1982} and was found to be consistent with two less precise measurements \cite{Detraz1972,Wilkinson1972}. The second largest source of uncertainty on $\delta$ stems from the half-life of $^{28}$P of 270.3(5) ms, which comes from a single measurement dating back to 1968 \cite{Armini1968} that was discounted as potentially inaccurate in early asymmetry parameter evaluations and more recent, but less precise measurements were used instead \cite{Wilkinson1972}. In turn, these various measurements conflicted with the previous evaluation value of 285(7) ms \cite{Endt1967}. It should be noted that a half-life of 285 ms for $^{28}$P would be sufficient to result in a null asymmetry parameter. The third largest source of uncertainty is the $Q_{\beta^+}$-value for a ground-state-to-ground-state decay transition of $^{28}$P. The $Q_{\beta^+}$ value in the AME20 \cite{AME2020}
comes from the $Q$-value measurement of the $^{28}$Si($^{3}$He,$t$)$^{28}$P reaction using the former Q3D magnetic spectrograph \cite{Wrede2010}. The precision and accuracy of the $Q_{\beta^+}$ value could easily be improved if it were measured directly using a Penning trap. 

\section{\label{sec:result}Conclusion}

Extensive searches for second-class currents based on the asymmetry parameter of $\beta^+$ and $\beta^-$ transitions feeding a common daughter nuclear state, mainly in the 1970s, have led to inconclusive results because of the large contribution from the uncertainty in various shell-model based nuclear structure corrections \cite{Wilkinson2000}. However, recent calculation results attempting to explain the peculiarly negative asymmetry parameter for the $^{28}$Al($\beta^-$)$^{28m}$Si and $^{28}$P($\beta^+$)$^{28m}$Si decays \cite{Xayavong2024}, drew our attention to the half-life of $^{28}$Al which was found to come from a series of conflicting measurements. Hence, a measurement of the half-life of $^{28}$Al was performed at the  Nuclear Science Laboratory \emph{TwinSol} facility and this new measurement gives a half-life of $^{28}$Al of 134.432(34) s. This new result is consistent with all the measurements plotted in Fig.~\ref{worldvalue} except for the 1978 measurement \cite{Becker1978}. This supports the disagreement with the 1978 value raised in \cite{ZAHN201370} and may warrant another independent measurement. If the 1978 measurement is excluded, a new world average for the $^{28}$Al half-life of 134.430(16)~s is obtained. With this value a slightly less negative asymmetry parameter of $-3.5(10)\%$  that still disagrees with the calculation results from \cite{Xayavong2024} is obtained. The main sources of uncertainties on the asymmetry parameter currently stem from the branching ratio (the dominant), the half-life, and the $Q_{\beta^+}$ value of the $^{28}$P $T=1$ transition. All of these come from single experiments; hence an independent confirmation at better precision is warranted. The $Q_{\beta^+}$-value can easily be improved by a measurement with a Penning trap. Similarly, both the branching ratio and half-life, which predate 1970, could easily be improved. These measurements will help to unambiguously confirm the negative value of the asymmetry parameter.

\section*{\label{sec:ac} Acknowledgments}

The authors would like to thank N.A.~Smirnova, L.~Xayavong, L~Hayen, N.~Severijns, and S.~Vanlangendonck for fruitful discussions. This work was supported by the National Science Foundation (NSF) Grants No. PHY-2310059, PHY-1401343 and PHY-1401242.

\nocite{*}

\bibliography{apssamp}

\end{document}